\documentclass[aps,prd,twocolumn,showpacs,preprintnumbers,amsmath,amssymb]{revtex4-1}
\usepackage{graphicx}
\usepackage[usenames,dvipsnames]{xcolor}
\usepackage{braket}
\usepackage{lineno}
\usepackage{comment}
\usepackage{amsthm}
\usepackage{mathtools}
\usepackage{bbold}
\usepackage[utf8]{inputenc}
\usepackage{subfigure}
\definecolor{winered}{rgb}{0.8,0,0}
\definecolor{darkb}{rgb}{0,0,0.8}
\usepackage[pdftex]{hyperref}
\hypersetup{
	colorlinks=true,
    linkcolor=winered,
    citecolor=blue,
    anchorcolor=black,
    menucolor=red,
    filecolor=magenta,
    urlcolor=darkb
}

\def\dmu{\partial_\mu}
\def\rar{\rightarrow}

\def\beq{\begin{equation}}
\def\enq{\end{equation}}
\def\bpl{\beta_{pl.}}
\def\bl{\beta_{l.}}

\graphicspath{{./figures/}}

\includecomment{versiona}

\DeclareMathOperator{\Tr}{Tr}

\begin{document}

\title{Discrete aspects of continuous  symmetries in the tensorial formulation of Abelian gauge theories}

\author{Yannick Meurice}
\email{yannick-meurice@uiowa.edu}
\affiliation{Department of Physics and Astronomy, 514 Van Allen Hall, The University of Iowa, Iowa City IA 52242}

\date{\today}

\begin{abstract}
We show that standard identities and theorems for lattice models with $U(1)$ symmetry get re-expressed discretely in the tensorial formulation of these models. 
We explain the geometrical analogy between the continuous  lattice equations of motion and the discrete selection rules of the tensors.
We construct a gauge-invariant transfer matrix  in arbitrary dimensions. We show the equivalence with its gauge-fixed version in a maximal temporal gauge and explain how 
a discrete Gauss's law is always enforced. We propose a noise-robust way to implement Gauss's law in arbitrary dimensions. 
We reformulate Noether's theorem for global, local, continuous or discrete Abelian symmetries: for each given symmetry, there is one corresponding tensor redundancy.
We discuss semi-classical approximations for classical solutions with periodic boundary conditions in two solvable cases. We show the correspondence of their weak coupling limit with the tensor formulation after Poisson summation. We briefly discuss connections with other approaches and implications for quantum computing. 
\end{abstract}
\maketitle
\section{Introduction}
Tensor Field Theory (TFT) is a recently developed approach of models studied in the context of lattice gauge theory
\cite{nishino96,Levin2006,Gu:2009dr,PhysRevLett.103.160601,Gu:2010yh,PhysRevB.86.045139,prb87,prd88,efratirmp,pre89,prd89,
PhysRevLett.115.180405,Shimizu:2014fsa,Shimizu:2014fsa,Takeda:2014vwa,Shimizu:2017onf,PhysRevB.98.235148,PhysRevLett.118.250602,Yoshimura:2017jpk,
Nakamura:2018enp,Kuramashi:2018mmi,
Kadoh:2018hqq,unmuth2018,Kadoh:2018tis,Kadoh:2018hqq,Bazavov:2019qih,Kadoh:2019ube,butt2019}. 
The basic idea is to rewrite the partition function of lattice models as a product of tensors where all the indices are contracted. 
Many lattice models have {\it compact} fields. This feature appears naturally when we  integrate over compact unitary groups attached to links in gauge models or over the Nambu-Goldstone modes of $O(N)$ symmetric scalar models in nonlinear sigma models. 
Functions over {\it compact} groups can be expanded in terms of {\it discrete} sums of characters for Abelian groups \cite{pdual} or more generally of group representations 
\cite{peter27}. This property was exploited to calculate strong coupling expansions \cite{balian75} and introduce dual variables \cite{fradkin78,kogut79,RevModPhys.52.453} for the type of lattice models mentioned above. 

These group theoretical methods were used in a systematic way to build the tensors \cite{prb87,prd88} of the spin and gauge models reviewed in \cite{kogut79}, and  to rewrite partition functions and averages of observables in a way that is suitable for exact coarse-graining or sampling of tensor configurations 
similar to the worm algorithm \cite{worm,PhysRevD.81.125007,pra90}.
It is also important to realize that TFT remains useful and accurate in regimes 
that are completely beyond the range of validity of the strong coupling expansion even when 
phase transitions are present \cite{foreman2018}. In addition, the discreteness of TFT formulations also makes them a natural starting point for building 
approximate forms of known lattice models suitable for quantum computations or quantum simulation experiments \cite{prd92,prl121}. 

Symmetry considerations have played a crucial role in uncovering the subconstituents of matter and their interactions. A key result is Noether's theorem which associates a conserved charge to a 
global {\it continuous} symmetry. Is there a way to re-express Noether's theorem in a completely 
{\it discrete} TFT formulation? In the following we will show show that the answer is affirmative in the case of a  continuous and compact $U(1)$ symmetry. We will also discuss the effect of approximations and various types of noise, which are unavoidable
in practical TFT implementations, on these symmetry properties. 

In the conventional formulation of field theory, a global $U(1)$ symmetry results in a conserved Noether current at the classical level. The use of the {\it continuous} symmetry and the classical equations of motion, which result from {\it local continuous variations} of the action, are crucial steps of the derivation. 
At the quantum level, the invariance under a local continuous shift of the field variables generates Schwinger-Dyson equations which are quantum versions of the equations of motion. For local 
$U(1)$ symmetries, Ward-Takahashi identities, or more complicated identities if gauge fixing is involved, can be found in quantum field theory textbooks such as Ref. \cite{peskin95}. 
These remarkable theorems and identities rely on the fact that the field variables are continuous. 

In this article we show that the basic features of continuous Abelian symmetries in the conventional formulation of field theory have discrete counterparts in TFT. The article is organized as follows. 
In Sec. \ref{sec:models}, we review the tensorial formulation of models with continuous Abelian symmetries in arbitrary Euclidean space-time dimension $D$. We start with the compact Abelian Higgs models (CAHM) and then consider the pure gauge limit and the O(2) spin model limit. 
In Sec. \ref{sec:eom}, we establish a precise correspondence between the classical equations of motion of the lattice action with the selection rules. 
They have identical grading and geometrical interpretation in terms of inside/outside features. 
In Ref. \cite{meurice2019}, it was noted that these selection rules, a discrete divergenceless  condition, could be interpreted as a discrete version of Noether's theorem and would also extend to discrete symmetries. The selection rules for the CAHM are a discrete 
version of Maxwell's equations with charges and currents. In particular, Gauss's law which has complicated aspects in the conventional Hamiltonian formulation appears in a transparent way in TFT. 
The questions of gauge-invariance and gauge-fixing are discussed in Sec. \ref{sec:gf}. 
We show that local selection rule redundancies observed in Ref. \cite{meurice2019} can be reinterpreted in terms of a gauge fixing that removes the integration over the fields leading to the redundant selection rules. In Sec. \ref{sec:noether}, we show that the mechanism can be extended to global symmetries and discrete symmetries. Noether's theorem for Abelian symmetries can be re-expressed in the tensor reformulation context as: for each symmetry, there is a corresponding tensor redundancy. This applies with a remarkable generality to local, global, continuous or discrete Abelian symmetries. 

The transfer matrix of the CAHM is constructed in arbitrary dimensions in Sec. \ref{sec:transfer}. 
It is made out of electric and magnetic ``layers" and is automatically gauge invariant. It defines a Hilbert space over which Gauss's law is implemented when we apply the transfer matrix on an arbitrary state. We use ``Gauss's law" in a context dependent manner. In the CAHM context, there are charges and currents and Gauss's law means  that the quantum numbers associated to the matter fields are completely fixed by the quantum numbers associated with the gauge fields. This is because there are infinitely many possible Fourier modes for bosonic fields in contrast to fermions that only allow a finite number of possibilities as, for instance,  for the Schwinger model \cite{Byrnes:2002nv,Banuls:2013jaa,Kuhn:2014rha,Buyens:2015tea,PhysRevD.98.074503,magnifico2019}.

However, when we take the pure gauge limit by decoupling the matter fields, we obtain a restriction on the gauge quantum number which is a discrete version of 
$\boldsymbol\nabla \cdot {\bf E}=0$. 
In the pure gauge limit, we discuss the equivalence with a gauge-fixed version and propose a way to implement Gauss's law with unrestricted variables which can be used in any dimension. In Sec. \ref{sec:ham}, we take the time-continuum limit in the same way 
as in \cite{kogut79} and get a similar Hamiltonian formulation. 

The correspondence between the continuous classical equations of motion and the discrete 
selection rules allows us to connect topological solutions that appear with periodic boundary conditions, and not with open boundary conditions, to tensor assemblies that are allowed or forbidden 
under the same periodic or open boundary conditions. In Sec. \ref{sec:top}, we show that this correspondence can be made precise using Poisson summation for two models that are exactly solvable. The practical consequences of the results for coarse graining, the continuum limit and quantum computations are briefly discussed in the conclusions.

\section{Abelian lattice models}
\label{sec:models}
In this section we introduce the compact Abelian Higgs model (CAHM) and situations where it can be reduced to the pure gauge $U(1)$ model or
the O(2) spin model. In its original form, the Abelian Higgs model has also a non-compact scalar field which can be decoupled by a strong coupling limit 
discussed in \cite{prd92} and will not be considered in the following. The main purpose of this section is to introduce models, notations and symmetries. 

In the following, we use a $D$-dimensional (hyper) cubic Euclidean space-time lattice. 
The space-time  sites are denoted $x=(x_1, x_2,\dots x_D)$, with $x_D=\tau$, the Euclidean time direction. 
Lattice units are implicit and the space-time sites are labelled with integers. We use the bold notation ${\bf x}$ for the $D-1$ dimensional labels of spatial sites. 
The links between two nearest neighbor lattice sites $x$ and $x+\hat{\mu}$ are labelled by $(x,\mu )$ and the plaquettes delimited 
by four sites $x$,  $x+\hat{\mu}$, $x+\hat{\mu}+\hat{\nu}$ and $x+\hat{\nu}$ are labelled by $(x,\mu,\nu)$. By convention, we start with the lowest index 
when introducing a circulation at the boundary of the plaquette. 
The total number of sites is denoted $V$. Periodic boundary conditions (PBC) or open boundary conditions (OBC) will be considered. 

Our main object is the CAHM partition function 
\beq
Z_{CAHM}=\prod_x\int_{-\pi}^{\pi}\frac{d\varphi _x}{2\pi}\prod_{x,\mu}\int_{-\pi}^{\pi}\frac{dA_{x,\mu}}{2\pi}
e^{-S_{gauge}-S_{matter}}, 
\label{eq:gaugemeasure}
\enq
with
\beq
S_{gauge}=\beta_{pl.}\sum_{x,\mu<\nu} (1-\cos(A_{x,\mu}+A_{x+\hat{\mu},\nu}-A_{x+\hat{\nu}, \mu}-A_{x,\nu})),
\label{eq:gauge}
\enq
and
\beq
\label{eq:smatter}
S_{matter}=\beta_{l.} \sum\limits_{x,\mu} (1-\cos(\varphi_{x+\hat{\mu}}-\varphi_x+A_{x,\mu})).\enq

The CAHM  is a gauged version of the O(2) model where the global symmetry under a $\varphi$ shift becomes local
\beq
\varphi_x'=\varphi_x+\alpha_x
\label{eq:varphix}
\enq
and these local changes in $S_{matter}$ are compensated by the gauge field changes
\beq
A_{x,\mu}'=A_{x,\mu}-(\alpha_{x+\hat{\mu}}-\alpha_x),
\enq
which also leave $S_{gauge}$ invariant. 

The matter fields can be decoupled by simply setting $\bl=0$. As they don't appear in the action, their integration 
yields a factor 1 and we are left with the pure gauge (PG) $U(1)$ lattice model with partition function
\beq
Z_{PG} =\prod_{x,\mu}\int_{-\pi}^{\pi}\frac{dA_{x,\mu}}{2\pi} e^{-S_{gauge}}.
\label{eq:gaugemeasure}
\enq

The decoupling of the gauge fields is less straightforward. Strictly speaking, the 
O(2) spin model is obtained by removing the gauge fields introduced to make the global symmetry a local one and the partition function of the 
O(2) model reads
\beq 
Z_{O(2)}=\prod_x\int_{-\pi}^{\pi}\frac{d\varphi _x}{2\pi}e^{-S_{O(2)}},
\enq
with 
\beq 
S_{O(2)}=\beta_{l.} \sum\limits_{x,\mu} (1-\cos(\varphi_{x+\hat{\mu}}-\varphi_x)).
\enq
It is tempting to consider this model as the weak gauge coupling limit 
($\bpl \rar \infty$) of the CAHM. However for compact gauge fields, this limit 
involves subtleties when defined in the context of the infinite volume and continuum limit.  
In addition, the O(2) model has charge sectors labelled by integers and it is possible to select a specific charge sector by tuning the gauge boundary conditions. This is discussed in Ref. \cite{prl121} in 1+1 dimensions. 

\def\zq{${\mathbb Z}_q\ $}
Most of the results presented in the rest of the paper also hold for finite subgroups of 
$U(1)$. If we consider the ``clock" restriction to angles $\varphi _x$ and $A_{x,\mu}$ taking values 
$\frac{2\pi}{q} \ell$ for $\ell=0,\ 1,\ \dots,\ q-1$, the values of $\ell$ are added modulo $q$ and form the additive group \zq.

\section{Tensor selection rules and lattice equations of motion}
\label{sec:eom}

In this section, we reformulate the CAHM in arbitrary dimensions using the tensor formalism developed in Refs. \cite{prd88,prd92}. 
We point out and explain the geometrical analogy between the tensor selection rules and the lattice equations of motion. We then discuss
the pure gauge and spin limits. 
\subsection{General case}
\label{subsec:general}

The basic ingredients of the tensor reformulation are the Fourier expansions for the links
 \begin{eqnarray}
           &\ & {\rm e}^{\bl  \cos(\varphi_{x+\hat{\mu}}-\varphi_x+A_{x,\mu})} = \cr
           &\ &\sum\limits_{n_{x,\mu}=-\infty}^{+\infty} {\rm e}^{i n_{x,\mu}(\varphi_{x+\hat{\mu}}-\varphi_x+A_{x,\mu})} I_{n_{x,\mu}}(\bl)\  , 
            \label{eq:foum}
\end{eqnarray}
and the plaquettes
 \begin{eqnarray}
   \label{eq:foug}
           &\ & {\rm e}^{\bpl  \cos(A_{x,\mu}+A_{x+\hat{\mu},\nu}-A_{x+\hat{\nu}, \mu}-A_{x,\nu})} = \\ \cr
           &\ &\sum\limits_{m_{x,\mu,\nu}=-\infty}^{+\infty} {\rm e}^{i m_{x,\mu,\nu}(A_{x,\mu}+A_{x+\hat{\mu},\nu}-A_{x+\hat{\nu}, \mu}-A_{x,\nu})} I_{m_{x,\mu,\nu}}(\bpl),\  \nonumber        
\end{eqnarray}
where the $I_n(\beta)$ are the modified Bessel functions of the first kind. 
Notice that in both cases, the argument of the cosine function and the expression multiplying the Fourier indices are identical and, in particular, their signs are identical. These signs can be interpreted as forming a binary grading. This grading depends on the link or plaquette to which the fields belong. 

The classical lattice equations of motions are obtained by setting the {\it derivatives} of the action with respect to the fields to zero. 
In general one obtains a sum of sines with relative signs corresponding to the grading. The tensors to be traced in the reformulation 
of the partition function, are obtained by {\it integrating} over the fields. When the Fourier indices corresponding to a given field are collected 
they appear with relative signs that correspond to the same grading and can be interpreted geometrically. We now discuss the scalar and gauge 
derivation/integration separately. 

For the scalar fields, we first introduce the notation 
\beq
d_{x,\mu}\equiv \varphi_{x+\hat{\mu}}-\varphi_x+A_{x,\mu}
\label{eq:covdev}
\enq
which approximates the covariant derivative of $\varphi$. 
The equation of motion
\begin{eqnarray}
\label{eq:eomo2}
\partial S/\partial \varphi_x&=&
\bl \sum_{\mu}[-\sin(d_{x,\mu})+\sin(d_{x-\hat{\mu},\mu})]\cr
&=&0.\end{eqnarray}
On the other hand the integration with respect to $\varphi_x$ implies
\beq
\sum_\mu[-n_{x,\mu}+n_{x-\hat{\mu},\mu}]=0.
\label{eq:noether}
\enq
It is clear that the geometrical structure of the two above equations are identical and that the 
equations can be obtained from each other by the substitution
\beq 
\label{eq:swap} \bl \sin(d_{x,\mu}) \leftrightarrow n_{x,\mu}.\enq
The geometrical interpretation is simple: in Eq. (\ref{eq:noether}), the $n_{x,\mu}$ come with a minus and correspond to links coming 
``out " of $x$ in the positive directions, while the $n_{x-\hat{\mu},\mu}$ come with a plus and correspond to links coming 
``in" the site  $x$ from the negative direction. Notice that this feature is completely dictated by the sign convention appearing 
in the Fourier expansion Eq. (\ref{eq:foum}). More specifically, fields appearing with a minus (plus) sign inside the cosine belong 
to an out (in) link, respectively. 

Notice that Eq. (\ref{eq:noether}) is a discrete version of Noether's theorem which is a divergenceless condition. This implies \cite{meurice2019} a discrete 
version of Gauss's theorem which is preserved when truncations are applied. In addition, the equations of motions are satisfied in average, when inserted in 
the path integral. This is a consequence of the invariance under a local shift for each integral which is used in the derivation of Schwinger-Dyson equations. 
If such a shift is applied after expansion in Fourier modes in the functional integral, then  Eq. (\ref{eq:noether}) follows, making the connection between the 
two sets of equations clear.

In a similar manner, we can assign in and out features to the plaquettes attached to a link in a way consistent with the Fourier expansion 
Eq. (\ref{eq:foug}). For $\mu<\nu$, $m_{x,\mu,\nu}$ are in and $m_{x-\hat{\nu},\mu,\nu}$ out,
 while for $\mu>\nu$, $m_{x,\nu ,\mu}$ are out and $m_{x-\hat{\nu},\nu,\mu}$ in. Using the obvious analogy with the continuum, we define the standard lattice field strength tensor
 \beq
 f_{x,\mu,\nu}\equiv A_{x,\mu}+A_{x+\hat{\mu},\nu}-A_{x+\hat{\nu}, \mu}-A_{x,\nu}.
 \enq
 As in the continuum they are gauge invariant. 
 
 With these notations,
\begin{eqnarray}
\label{eq:contmax}
\partial S/\partial A_{x,\mu}&=&\bpl \sum_{\nu>\mu}[ \sin(f_{x,\mu,\nu})-\sin(f_{x-\hat{\nu},\mu,\nu})]\cr
&&+\bpl \sum_{\nu<\mu}[-\sin(f_{x,\nu,\mu})+\sin(f_{x-\hat{\nu},\nu,\mu})]\cr
&&+\bl \sin(d_{x,\mu})\cr
&=&0.
\end{eqnarray}
On the other hand, the integration over $A_{x,\mu}$ yields the selection rule
\begin{eqnarray}
\label{eq:discrmax}
&\ & \sum_{\nu>\mu}[m_{x,\mu,\nu}-m_{x-\hat{\nu},\mu,\nu}]\cr
&+&\sum_{\nu<\mu}[ -m_{x,\nu,\mu}+m_{x-\hat{\nu},\nu,\mu}]\cr
&+&n_{x,\mu}\cr
&=&0.
\end{eqnarray}
We see again geometric similarities between Eq. (\ref{eq:contmax})  with continuous variable and Eq. (\ref{eq:discrmax}) with integer variables. 
They both have the tensor structure of \beq \partial _\mu F^{\mu \nu}=J^\nu.\enq
They can be mapped into each other using Eq. (\ref{eq:swap}) and 
in addition the substitution 
\beq
\bpl  \sin(f_{x,\mu,\nu}) \leftrightarrow m_{x,\mu,\nu}.
\enq

Eq. (\ref{eq:discrmax}) means that the link indices $n_{x,\mu}$ can be seen as determined by unrestricted plaquette indices $m_{x,\mu,\nu}$. 
We write this dependence as $n_{x,\mu}(\{m\})$ which is shorthand for Eq. (\ref{eq:discrmax}). Note that for $n_{x,\mu}(\{m\})$
Eq. (\ref{eq:noether}) holds \cite{meurice2019} and as long as the gauge fields are present, there is no need to enforce Eq. (\ref{eq:noether}).

Each integration provides a tensor with the selection rules discussed above. For convenience we factorize all the $I_0(\beta)$ factors which dominate the 
small $\beta$ regime and define the ratios 
\beq 
\label{eq:tn}
t_n (\beta) \equiv \frac{I_n(\beta)}{I_0(\beta)} \simeq\begin{cases} 1-\frac{n^2}{2\beta}+ \mathcal{O}(1/\beta^2),\  {\rm for}\  \beta \rar \infty \\
\frac{\beta^n}{2^n n!}+ \mathcal{O}(\beta^{n+2}), \  {\rm for}\ \beta \rar 0 \end{cases}.\enq
Their limiting behavior at weak and strong coupling will be used often.

\begin{figure}[h]
 \centering
 \includegraphics[width=6cm]{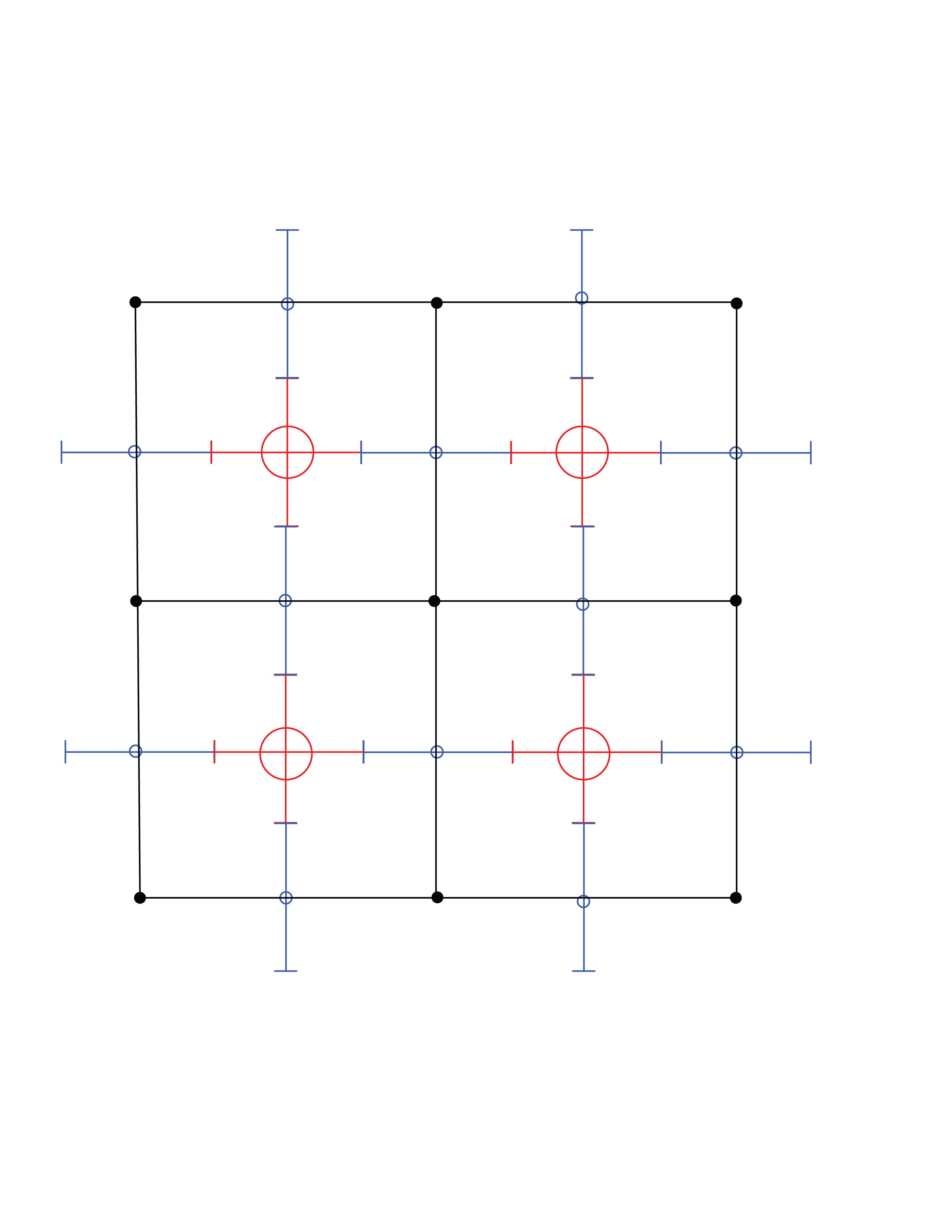}
\caption{\label{fig:AandB}$A$ and $B$ tensors assembled  in $D=2$.
Small circles (blue online) are used for the $A$- tensors  and
large circles  (red online) for the $B$-tensors.} 
\end{figure}

The four tensor legs attached to a given plaquette $(x,\mu,\nu)$ must carry the same index $m$. For this purpose we introduce the ``$B$-tensor" as in \cite{prd88}
         \beq
            \label{eq:ttensorB}
                B^{(x,\mu,\nu)}_{m_1m_2m_3m_4}
            =\begin{cases} 
                t_{m_1}(\bpl ),  &\mbox{if all } m_i\mbox{ are the same} \\
                0, & \mbox{otherwise}.
            \end{cases}
        \enq
These are assembled (traced) together with ``$A$-tensors" attached to links with $2(D-1)$ legs orthogonal to the 
link
  \beq
            \label{eq:Atensor}
                       A^{(x,\mu)} _{m_1\dots m_{2(D-1)}}=t_{n_{x,\mu}}(\bl)
           \delta_{n_{x,\mu},n_{x,\mu}(\{m\})}.
        \enq
Notice that in contrast to Ref. \cite{prd88}, the weight of the plaquettes is carried by the $B$-tensor. 
The partition function with PBC can now be written as
        \begin{eqnarray}
        \label{eq:traceall}
            Z&=&(e^{-\bpl} I_0(\bpl))^{VD(D-1)/2}(e^{-\bl} I_0(\bl))^{VD}  \cr
            &&\cr
            &\times &\Tr \prod_{l.}A^{(l.)}_{m_1, \dots m_{2(D-1)}}\prod_{pl.}B^{(pl.)}_{m_1m_2m_3m_4},
      \end{eqnarray}            
where the trace means index contraction following the geometric procedure described above. 
The tensor assembly is illustrated in Fig. \ref{fig:AandB} for $D=2$. Illustrations for $D=3$ will be provided in Sec. \ref{sec:transfer}.

It is clear that for PBC, we have a discrete translation invariance and the tensor assembly is the same everywhere. 
We can introduce  OBC by starting with PBC and setting $\bl$ and $\bpl$ to zero on the links and plaquettes at the boundary.
Since 
\beq
I_n(0)=\delta_{n,0},
\label{eq:betanot}
\enq
this forces the indices at the boundary to be zero with an associated weight 1.

\subsection{The O(2) model}

Part of the results of subsection \ref{subsec:general} extend in a straightforward way to the O(2) model.
We just need to set $A_{x,\mu}=0$ in Eqs. (\ref{eq:covdev}) to (\ref{eq:noether}). It was pointed out \cite{meurice2019}, 
that Eq. (\ref{eq:noether}) is a discrete version of Noether's theorem associated with the global O(2) symmetry. 
A discrete version of Gauss theorem holds and guarantees a global neutrality for PBC and OBC.

It is also interesting to keep the gauge fields and take the $\bpl \rar \infty$ limit where the weights of the $B$-tensors are 1. 
The relation $n_{x,\mu}(\{m\})$ of Eq. (\ref{eq:discrmax}) remains valid and guarantees that the discrete divergenceless 
condition Eq.  (\ref{eq:noether}) is obeyed for arbitrary plaquette configurations $\{m\}$.
For $D=2$, this corresponds to the dual construction \cite{RevModPhys.52.453}, 
however, the gauge procedure described here extends to any dimension without requiring the use of the dimension-dependent 
Levy-Civita tensor.

\subsection{Pure gauge limit}
\label{subsec:pure}
We now consider the pure gauge limit by setting $\bl =0$. Eq. (\ref{eq:betanot}) imposes the 
constraint $n_{x,\mu}=0$ and Eq. (\ref{eq:discrmax}) reduces to a discrete version of $\partial _\mu F^{\mu \nu}=0$. 
We can make this statement more precise by introducing suggestive notations. 
We define the electric integers 
\beq e_{x,j}\equiv m_{x,j,D},\enq
with $j=1,\dots,D-1$, the integers associated with time plaquette and which can be interpreted 
as electric fields.  Eq. (\ref{eq:discrmax}) for $\mu =D$ reads
\beq
\sum_{j=1}^{D-1}(e_{x,j}- e_{x-\hat{j},j})=0.
\enq
This is a discrete form of Gauss's law in the pure gauge limit $\boldsymbol\nabla \cdot {\bf E}=0$.

For $D\geq 3$, we can introduce magnetic fields in a dimension dependent way. 
For $D=3$, we define
\beq
b_x\equiv m_{x,1,2}.
\enq
Eq. (\ref{eq:discrmax}) for $\mu =1$ and 2 are
\begin{eqnarray}
 e_{x,1}- e_{x-\hat{\tau},1}&=&-( b_{x}- b_{x-\hat{2}}),\cr
 e_{x,2}- e_{x-\hat{\tau},2}&=&( b_{x}- b_{x-\hat{1}}).
\end{eqnarray}
These are a discrete version of the $D=3$ Euclidean Maxwell's equations 
\begin{eqnarray}
\partial _1 B&=&\partial_\tau E_2\cr
\partial _2 B&=&-\partial_\tau E_1,
\end{eqnarray}
with $B=F^{12}$. 
However, there is {\it no} discrete equation corresponding to Maxwell equation for the dual field strength tensor
\beq
\label{eq:3dual}
\dmu\epsilon^{\mu \nu\sigma}F_{\nu \sigma}=0.
\enq
Example of legal configurations violating the discrete version of Eq. (\ref{eq:3dual}), also written $\dot{B}=-\boldsymbol\nabla \times {\bf E}$, can be constructed. 

For $D=4$, we can introduce 
\beq
 b_{x,j}\equiv \epsilon _{jkl}m_{x,k,l},
\enq
and obtain a discrete version of 
\beq
\label{eq:max2}
\partial_\tau{\bf E}=-\boldsymbol\nabla \times  {\bf B}, 
\enq
with the Euclidean magnetic field
\beq
F^{jk}=+\epsilon^{jkl}B^l. 
\enq
Note that Eq. (\ref{eq:max2}) implies 
\beq
\label{eq:dgauss}
\partial_\tau(\boldsymbol\nabla \cdot {\bf E})=0,
\enq
even if we don't impose Gauss's law.
Again there is no discrete version of the homogeneous equations for the dual field strength 
 $\dot{\bf  B}=-\boldsymbol\nabla \times {\bf E}$  and $\boldsymbol\nabla \cdot  {\bf B}=0$. Note that 
 the sign in Eq. (\ref{eq:max2}) is different in Euclidean and Minkowskian spaces. It can be traced to the 
 minus sign in the Minkowskian Klein-Gordon equation. 
 
 \subsection{Restrictions to \zq} 
Some of the results of this section hold in an obvious way for the \zq  restrictions. The infinite sums in the Fourier expansions are replaced by finite sums with $q$ values. The modified Bessel functions are 
replaced by their discrete counterparts:
\beq
I_n(\beta) \rar I_n^{(q)}(\beta)\equiv (1/q)\sum_{\ell=0}^{q-1}e^{\beta \cos(\frac{2\pi}{q} \ell)} e^{i n\frac{2\pi}{q} \ell},\enq
which in the large $q$ limit turns into the usual integral formula. In the Ising case ($q=2$), we have 
\beq
I_0(\beta) \rar \cosh(\beta), \ {\rm and} \ I_1(\beta) \rar \sinh(\beta).
\enq
The selection rules 
Eqs. (\ref{eq:noether}) and (\ref{eq:discrmax}) remain valid modulo $q$. The infinitesimal variations of the action can be replaced discrete variations by an amount $\frac{2\pi}{q}$, but the sine functions  should be replaced by finite differences of cosine functions. 

 \section{Local gauge invariance, selection rule redundancy and gauge fixing}
 \label{sec:gf}
 
 Sec. \ref{subsec:general} makes clear that selection rule Eq. (\ref{eq:noether}) due to the $\varphi$ integration is redundant and 
 a consequence of Eq. (\ref{eq:discrmax}). This forces a {\it local neutrality}. If we insert $e^{i\varphi_x}$ in the partition function 
 Eq. (\ref{eq:noether}) is modified and clashes with the original form of Eq. (\ref{eq:noether}) which follows from Eq. (\ref{eq:discrmax}),  forcing the functional integral to be zero. 

\begin{figure}[h]
 \centering
\includegraphics[width=5cm]{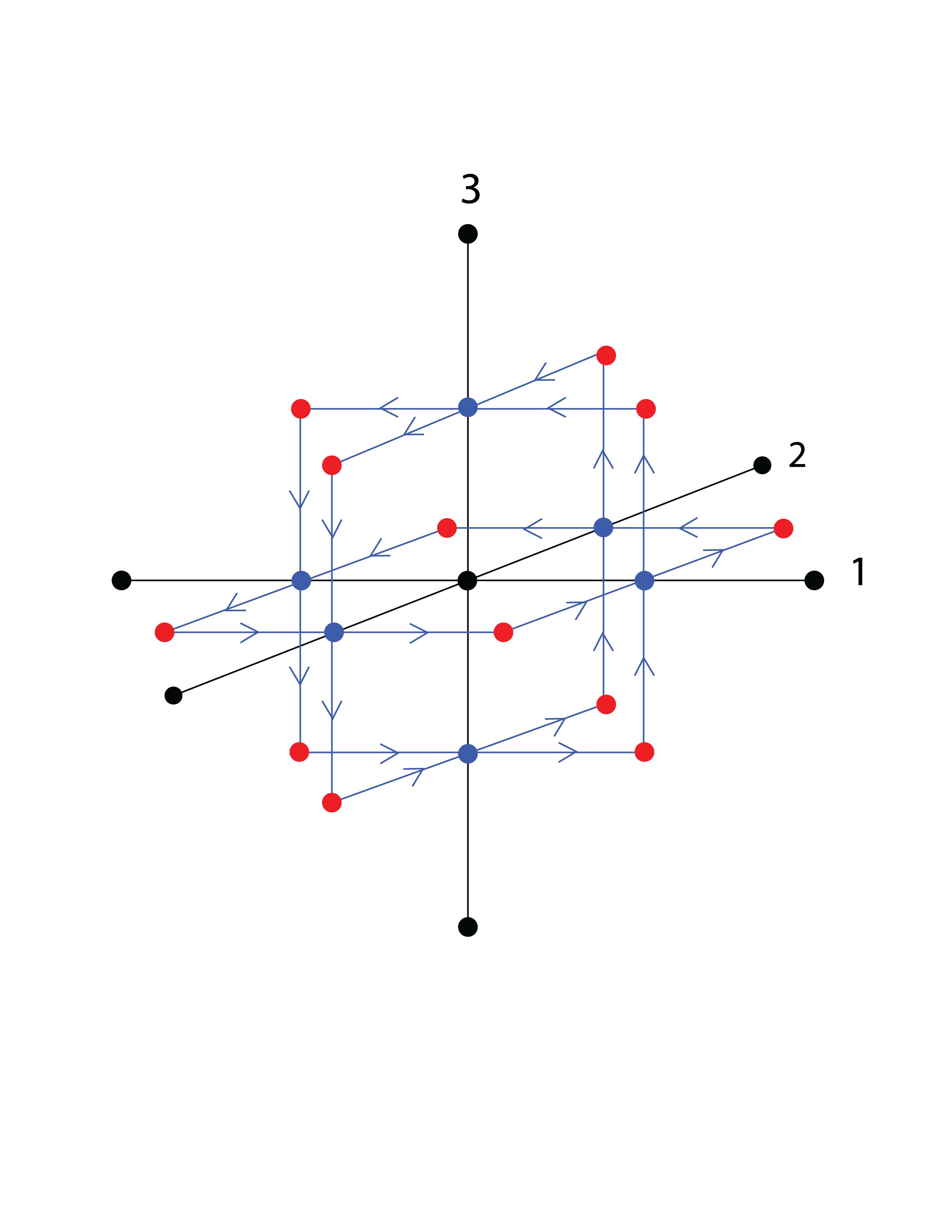}
\caption{ \label{fig:red}Illustration that one divergenceless condition  is redundant for $D$= 3. 
Imagine the tensor assembled on the surface of a cube, remove the $A$-tensor on the top: the sum of the in indices equals the sum of the out indices at the missing tensor because it holds at the 17 other vertices.}
\end{figure}

Similarly, it  was shown \cite{meurice2019} that in the pure gauge limit the set of equations (\ref{eq:discrmax}) with $n_{x,\mu}=0$ are not independent. 
If we pick a site, we can construct a in-out partition for the legs attached to links coming out of this site, the sum of  ``in" indices is the same as the sum of the ``out" indices, and if we assemble them on the boundary of a $D$-dimensional cube,  as illustrated in 
Fig. \ref{fig:red} for $D$= 3, one of the divergenceless condition follow from the other $2D-1$ conditions. 

To be completely specific, we review the details of this in-out partition \cite{meurice2019}.
For a given pair of directions $\mu$ and $\nu$, there are 8 types of legs for the $A$-tensors on links connected the site $x$ that we label $[(x,\mu),\pm \hat{\nu}]$, $[(x-\hat{\mu},\mu),\pm \hat{\nu}]$, $[(x,\nu ),\pm \hat{\mu}]$, and $[(x-\hat{\nu},\nu),\pm \hat{\mu}]$. 
The pair of indices appearing first refers to the links where the $A$-tensor is attached and the second index to the direction of the leg which can be positive or negative. The $[(x,\mu), \hat{\nu}]$ with $\mu<\nu$ are given an out assignment. There are three operations that swap in and out: changing $(x,\mu)$ into $(x-\hat{\mu},\mu)$, changing $\hat{\mu}$ into $-\hat{\mu}$ and interchanging $\mu$ and $\nu$.

 This redundancy can be rephrased in a more enlightening way in the 
 discrete electric/magnetic language developed in Sec. \ref{subsec:pure}: if Gauss's law is satisfied for a $A$-tensor attached to the $(({\bf x},\tau), D)$ time link which is assembled with 
 the divergenceless $A$-tensors attached to the $2(D-1)$ spatial links $(({\bf x},\tau +1), j)$ and $(({\bf x}-\hat{j},\tau +1), j)$ with $j=1,\dots, D-1$, then the  
 $A$-tensors attached to the time link $(({\bf x},\tau+1), D)$ is forced to obey Gauss's law because of its connection to the other tensors. 

We can now see that gauge fixing is equivalent to removing these redundant conditions. 
For the CAHM, we can go to the unitary gauge where the $\varphi$ can be removed everywhere and the 
redundant Eq. (\ref{eq:noether}) obtained from the $\varphi$ integration disappears independently of boundary conditions. For the pure gauge case, 
we can try to use the temporal gauge to set $A_{x,D}$ to zero. For OBC, this can be accomplished for all time links. 
From Eq. (\ref{eq:betanot}), OBC imply $e_{x,i}=0$ on the two time slices at the boundary (one below the initial time and one above the final time). In other words, it corresponds to a transition from the state where there is no electric field into 
itself. 
As Gauss's law is satisfied by the trivial configuration at the time slices at the boundaries it is also satisfied 
on every time slice. 
For PBC, one link remains to be integrated for each closed time line (Polyakov loop) attached to any given spatial site. 
Putting these unintegrated time links at the same time, we get a time layer where after integrating over the leftover time links, 
Gauss's law is satisfied. Again, Gauss's law is then propagated to the entire lattice for the reason explained above. 

\section{A reformulation of Noether's theorem}
\label{sec:noether}

The discussion of Sec. \ref{sec:gf} clarifies that redundant selection rules are in one-to-one correspondence with irrelevant integrations. 
In other words, we can skip the integrations that produce redundant selection rules and replace these integrated fields by arbitrary values.
This is exactly what gauge-fixing does. 
With our normalization of each integration over the circle to one, this does not cost extra factors. 

The argument can be extended to global symmetries. In the case of the O(2) model, it follows from the discussion of Ref.  \cite{meurice2019} that in-out assignments for the 
$2D$ legs of the divergenceless tensor attached to sites imply that {\it one} of the divergenceless conditions is a consequence of all the other ones. This requires the 
whole tensor network to be isolated. For PBC, there are no boundaries. For OBC, the boundaries carry 0 indices which are neutral (neither in or out). 
It is clear that the global O(2) symmetry allows us to fix {\it one} of the $\varphi$ fields to an arbitrary value. 

The redundancy argument extends to discrete \zq subgroups of $U(1)$ where the divergenceless condition is expressed modulo $q$ and the infinite set of Bessel functions are 
replaced by the $q$ discrete ones. 

In view of this discussion, we suggest that Noether's theorem can be expressed in the tensor formulation context as: for each symmetry, there is a corresponding tensor redundancy. This applies to global, local, continuous and discrete Abelian symmetries. 

\section{Transfer matrix}
\label{sec:transfer}
In Eq. (\ref{eq:traceall}),  the partition function is written as the trace of a product of tensors attached to links and plaquettes. We can organize this trace by assembling ``time layers"  corresponding to ``magnetic" time slices and  ``electric" slices half-way between the magnetic time slices. This construction singles out a time direction as for the Hamiltonian treatment. 
The case $D=2$ is discussed in Ref. \cite{prd92} and the pure gauge $D=3$ case in Ref. \cite{unmuth2018}.  
For $D=3$, this construction can be visualized as a ``lasagna". 
We first discuss the general CAHM case and then the two limits. 
\subsection{General case}

\begin{figure}[h]
 \centering
 \vskip-60pt
\includegraphics[width=8cm]{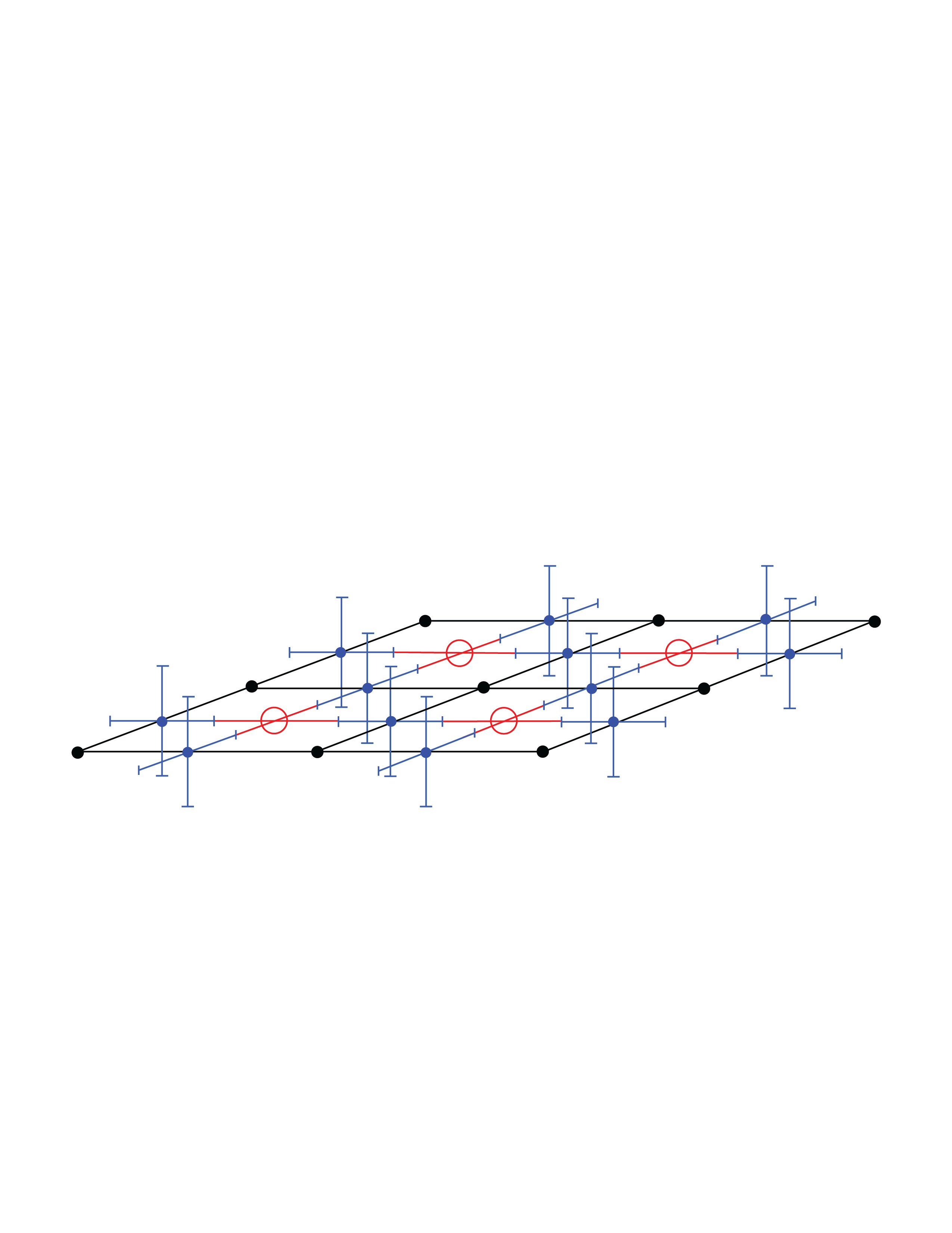}
 \vskip-40pt
\caption{\label{fig:mlayer}Magnetic layer of the transfer matrix for $D=3$ on a time slice. 
Small circles (blue online) are used for the $A$- tensors  and
large circles  (red online) for the $B$-tensors.
} 
\end{figure}

For the CAHM, the magnetic time slices contain $B$-tensors 
on space-space plaquettes and the $A$-tensors attached to their space links. These $A$-tensors 
have $2(D-2)$ legs in spatial directions and 2 legs in opposite time directions which we can visualize as the ``past" and the ``future".  These legs 
in the time direction are connected to space-time plaquettes. This is illustrated in Fig. \ref{fig:mlayer} for $D=3$. Seen ``from above", in other words without the time legs, 
this looks like the full $D=2$ assembly shown in Fig. \ref{fig:AandB}

\begin{figure}[h]
 \centering
  \vskip-40pt
\includegraphics[width=6cm]{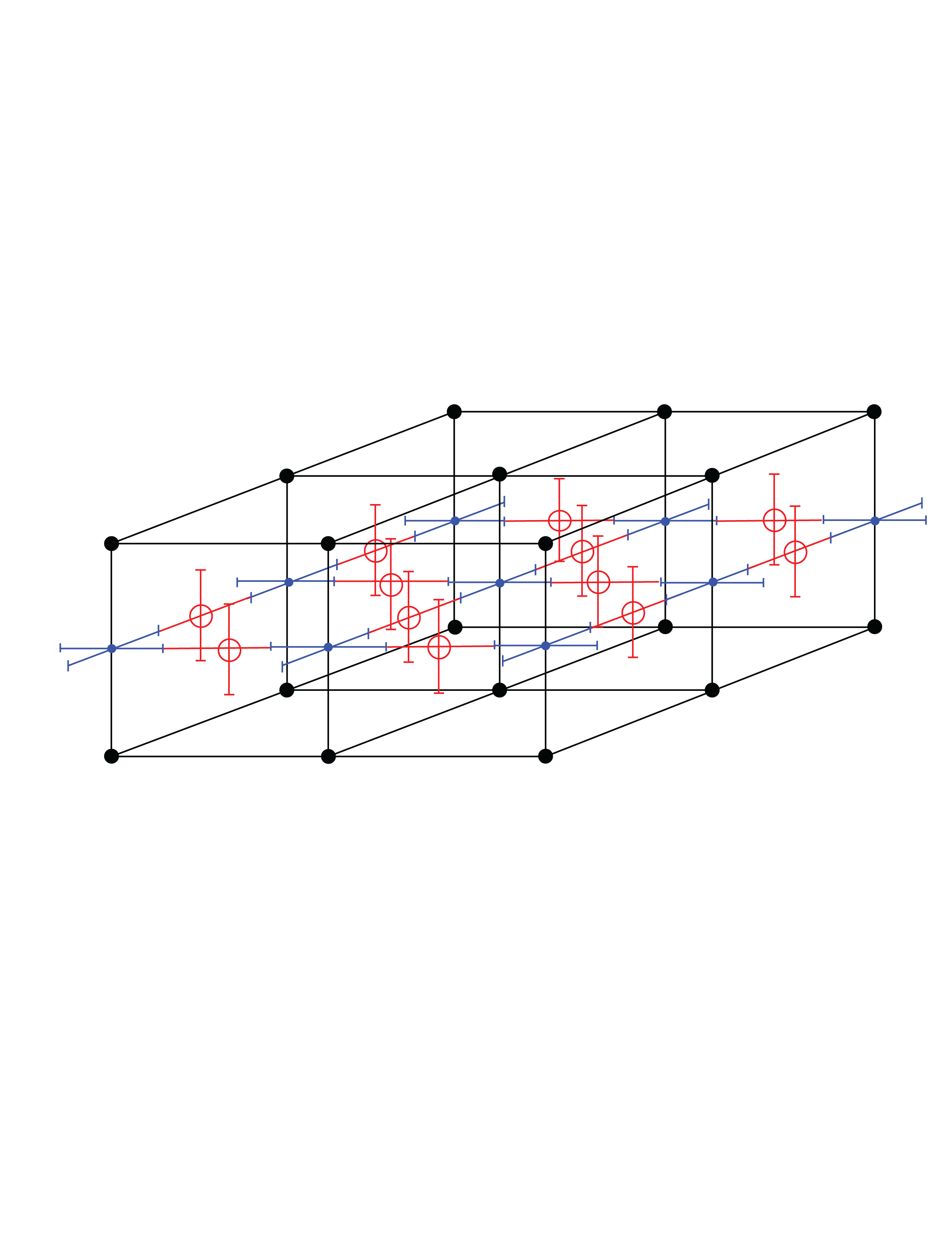}
\vskip-60pt
\includegraphics[width=6cm]{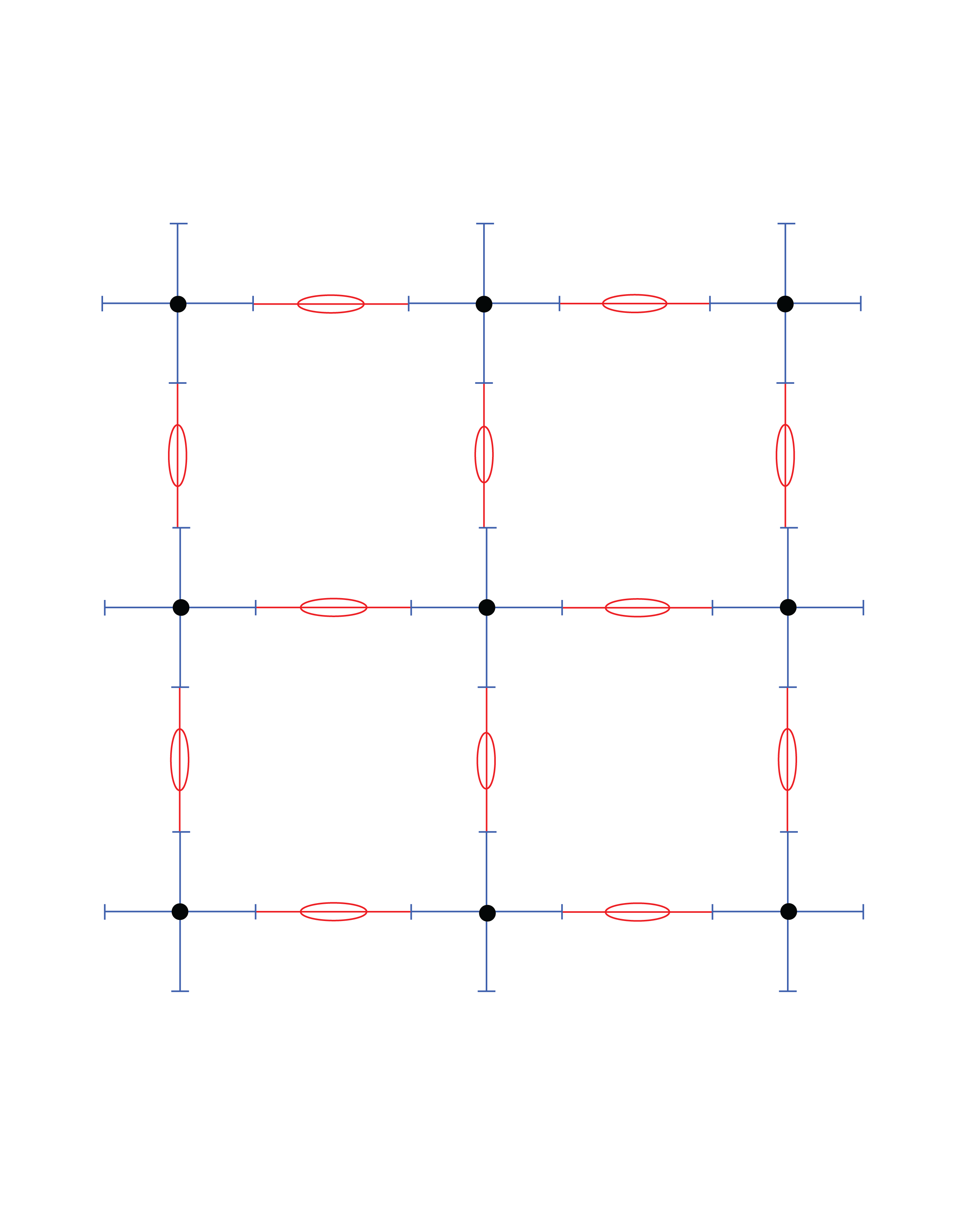}
 \vskip-40pt
\caption{\label{fig:elayer}Electric layer of the transfer matrix for $D=3$ between two time slices (top),  small circles (blue online) are used for the $A$- tensors  and
large circles  (red online) for the $B$-tensors, and ``from above" (bottom).} 
\end{figure}

In between the magnetic time slices we have electric layers with $B$-tensors 
on space-time plaquettes labelled by $ e_{({\bf x},\tau),j}$ with a fixed $\tau$, and the $A$-tensors attached to their time links. These $A$-tensors 
have $2(D-1)$ legs all in spatial directions. This is illustrated in Fig. \ref{fig:elayer} for $D=3$. Seen ``from above", in other words without the time legs of the $B$-tensors, 
this looks like the full $D=2$ assembly for the O(2) model. 

We want to represent these two types of layers as matrices. 
It is convenient to think of these two types of layers as matrices connecting electric states
\beq
\label{eq:hilbert}
\ket{\{ {\bf e} \} }=\otimes_{{\bf x},j}\ket{ e_{{\bf x},j}}. 
\enq
This is a natural choice because the $B$-tensors on the space-time plaquettes have two legs in the 
time direction. 
In this basis, the electric layer can be expressed as a diagonal matrix ${ \mathbb T}_E$ with 
matrix elements
\beq
\bra{\{ {\bf e'} \} } { \mathbb T}_E\ket{\{ {\bf e} \} }=\delta_{\{ {\bf e} \} ,\{ {\bf e'} \} } T_E(\{ {\bf e} \}), 
\enq
where  $T_E(\{ {\bf e} \})$ can be written with some implicit notations as a traced product of $A$ tensors on time links with 
$B$ tensors on space-time plaquettes
\beq
T_E(\{ {\bf e} \})=\Tr \prod_{time\ l.}A^{(l.)}_{m_1, \dots m_{2(D-1)}}\prod_{sp.-time\ pl.}B^{(pl.)} ({\bf e}).
\enq
Similarly, we can define a magnetic  matrix ${ \mathbb T}_M$ with 
matrix elements
$\bra{\{ {\bf e} \} } { \mathbb T}_M\ket{\{ {\bf e'} \} }$ with the indices ${\bf e}$ and ${\bf e'}$ carried by the time legs of the $A$-tensors. 
\begin{eqnarray}
&&\bra{\{ {\bf e'} \} } { \mathbb T}_M\ket{\{ {\bf e} \} }=\cr
&&\Tr \prod_{sp.\ l.}A^{(l.)}_{m_1, \dots m_{2(D-1)}}({\bf e},{\bf e'})\prod_{sp.-sp.\ pl.}B^{(pl.)}.
\end{eqnarray}
All the traces are over the spatial legs of the tensors, while the time legs are left open and carry the the indices ${\bf e}$ and ${\bf e'}$. 
Figs. \ref{fig:mlayer} and \ref{fig:elayer} should help visualizing these matrix elements: the horizontal lines correspond to traced indices while 
the vertical indices carry the $\{ {\bf e} \} $ indices. 

We can now define the transfer matrix ${ \mathbb T}$ as  
\begin{eqnarray}
        \label{eq:transfer}
            { \mathbb T}&\equiv&(e^{-\bpl} I_0(\bpl))^{(V/N_\tau)D(D-1)/2}(e^{-\bl} I_0(\bl))^{(V/N_\tau)D}  \cr
            & &\times { \mathbb T}_E^{1/2}{ \mathbb T}_M{ \mathbb T}_E^{1/2},
      \end{eqnarray}    
with $N_\tau$ the number of sites in the temporal direction. 
With this definition, we can reexpress the partition function as 
  \beq    
   Z=   \Tr { \mathbb T}^{N_\tau},
      \enq
\subsection{O(2) limit}
\label{subsec:to2}
For the O(2) model, the transfer matrix can be constructed by taking all the O(2) tensors on a time slice and tracing over the spatial indices. This is discussed in detail in 
Ref. \cite{pra90} in the case $D=2$ and the extension to arbitrary dimension is straightforward. In the CAHM reformulation 
the implicit O(2) tensors attached at each site are divergenceless. By implicit, we mean that
a link index $n_{x,\mu}$ carries a weight $t_{n_{x,\mu}}(\bl )$ as in the O(2) model (see Eq. (\ref{eq:Atensor})). 

From the perspective of quantum simulations, the gauge  parametrization insures that the divergenceless condition is automatically satisfied and insensitive to noise.
However the Hilbert space becomes larger for $D\ge 4$ (see below).

\subsection{Pure gauge limit: a robust way to implement Gauss's law}
\label{subsec:robust}

The restricted electric Hilbert space of the pure gauge compact $U(1)$ model in $D$ dimensions is equivalent to the set of legal tensor configurations of a $D-1$ dimensional O(2) model. The integer quantum numbers of the 
space-time plaquettes $e_{x,j}$ are like the link variables $n_{x,j}$ for O(2). Both sets are divergenceless. For the pure gauge model, the divergenceless condition is Gauss's law 
(without charge density). 

In Sec. \ref{subsec:to2}
we presented  the O(2) model as a weak gauge coupling limit of the compact Abelian Higgs model. As seen in Eq. (\ref{eq:discrmax}),  integration over the gauge fields 
provides an automatic divergenceless condition for the link quantum numbers. This is a discrete version of $\dmu \partial_\nu F^{\mu \nu}=\dmu J^\mu= 0$. The argument does not involve the dimensional-dependent Levy-Civita tensor. 

We can insure that Gauss's law is automatically satisfied by introducing a new set of quantum numbers 
$ c_{{\bf x},j,k}$, associated with the plaquettes of a $D-1$ CAHM, and unrelated to the existing gauge quantum numbers. For an arbitrary configuration $\{  c_{{\bf x},j,k}  \} $, we impose
\begin{eqnarray}
\label{eq:discrmax2}
e_{x,j}&=&\sum_{k>j}[- c_{x,j,k}+ c_{x-\hat{k},j,k}]\cr
&+&\sum_{k<j}[  c_{x,k,j}- c_{x-\hat{k},k,j}],
\end{eqnarray}
and Gauss's law is automatically satisfied. 
This is a discrete version of 
\beq
E^k=\partial _j C^{jk},
\enq
For an arbitrary antisymmetric tensor $C^{jk}$ with indices $j,\ k$ running from 1 to $D-1$. 
It is possible to introduce dimension-dependent ``magnetic" notations such as  $G=\epsilon^{kl}C^{kl}$ for $D=3$ and $G^j=\epsilon^{jkl}C^{kl}$ for $D=4$.

For a $D=3$ pure gauge theory we can visualize the electric Hilbert space as a $D=2$ O(2) model 
being on a plane between two time slices, as at the bottom of Fig. \ref{fig:elayer}. We can further imagine the auxilliary variables located in the middle of the plaquettes of this ``horizontal" plane, which means 
in the center of the $D=3$ cubes of the original lattice.
This is equivalent to the dual formulation discussed in Ref. 
\cite{unmuth2018}.

For $D=4$, this reparametrization is a discrete equivalent of 
setting 
\beq
{\bf E}=\boldsymbol\nabla \times {\bf G}.
\enq
This guarantees Gauss's law, but $\boldsymbol\nabla \times {\bf E}$ is in general non-zero  so we don't use this trick for conventional electrostatics because one of the homogeneous Maxwell's equation
($\dot{\bf B}=-\boldsymbol\nabla \times {\bf E}$) implies
that the magnetic field changes with time. 

This method is very efficient for $D=3$, because it reduces the dimensionality of the Hilbert space. 
There is one index per site ($ c_{x,1,2}$) rather than 2 ($ e_{{\bf x},1}$ and $ e_{{\bf x},2}$). For $D=4$, there are 3 indices per sites in both case, because $ c_{x,j,k}$ is only defined up to a gradient. However, the robustness against noise is an important advantage. At the end of Sec. \ref{sec:ham}, we briefly discuss possible optimizations.

\section{Hamiltonian limit}
\label{sec:ham}

For lattice models at Euclidean time, the transition from the Lagrangian to the Hamiltonian formalism is a standard procedure \cite{fradkin78,kogut79}. 
The central idea is to deform the isotropic formulation by increasing the $\beta$ variables associated with time directions and to decrease those associated with space directions.
In the tensor language, examples involving the transfer matrix in 1+1 dimensions \cite{pra90,prd92,prl121} and  2+1 dimensions \cite{judah.thesis,unmuth2018} provide steps that will be followed below. 

\def\te{${ \mathbb T}_E\ $}
\def\tm{${ \mathbb T}_M\ $}

The crucial features of \te is that it only involves time links and plaquettes having one direction in time. We introduce separate $\beta_\tau$ couplings for \te and use redefinitions in terms of the time lattice spacing $a_\tau$:
\beq
\beta _{\tau pl.} =\frac{1}{a_\tau g^2_{pl.}}, \ {\rm and}\  \beta _{\tau l.}=\frac{1}{a_\tau g^2_{l.}}.\enq
Given the weak coupling (large  $\beta$) behavior of $t_n(\beta)$ given in Eq. (\ref{eq:tn}), 
at first order in $a_\tau$, we get ``rotor" energies $(1/2)g^2_{pl.} m^2$ for the plaquettes and 
$(1/2)g^2_{l.} n^2$ for the links.

On the other hand, \tm only involves space links and space-space plaquettes and we redefine 
\beq
\label{eq:contlim}
\beta _{s\  pl.} =a_\tau J_{pl.}, \ {\rm and}\  \beta _{s\  l.}=a_\tau h_{l.}.\enq
Given the strong coupling (small $\beta$) behavior of $t_n(\beta)$ from  Eq. (\ref{eq:tn}), 
at first order in $a_\tau$, the contribution to \tm involve a {\it single} link or plaquette with 
quantum number $\pm 1$ all the other ones having a quantum number 0 and a weight 1. This leaves us with only few options:
raise or lower $ e_{{\bf x},j}$ over a link $({\bf x},j)$ or raise over two links and lower over the two other links of a plaquette.

We define the Hamiltonian ${\mathbb H}$ as the order $a_\tau$ correction to the identity 
in the transfer matrix:
\beq 
{ \mathbb T}=\mathbb{1}-a_\tau {\mathbb H}+\mathcal{O}(a_\tau ^2)
\enq
After introducing the operators \cite{banks76}
$\hat e_{{\bf x},j}$ and $\hat U_{{\bf x},j}$ such that 
\begin{eqnarray}
\hat e_{{\bf x},j}\ket{ e_{{\bf x},j}}&=& e_{{\bf x},j}\ket{ e_{{\bf x},j}}\cr
\hat U_{{\bf x},j}\ket{ e_{{\bf x},j}}&=&\ket{ e_{{\bf x},j}+1}\\ 
\hat U^\dagger _{{\bf x},j}\ket{ e_{{\bf x},j}}&=&\ket{ e_{{\bf x},j}-1},\cr\nonumber
\end{eqnarray}
the discussion of the first order behavior of \te and \tm allows us to write
\begin{eqnarray}
\label{eq:ksham}
 {\mathbb H}=&&\frac{1}{2}g^2_{pl.} \sum_{{\bf x},j}(\hat e_{{\bf x},j})^2\cr
 &+& \frac{1}{2}g^2_{l.}(\sum_{{\bf x},j}(\hat e_{{\bf x},j}-\hat e_{{\bf x}-\hat{j},j}))^2\cr
 &-& h_{l.} \sum_{{\bf x},j}(\hat{ U}_{{\bf x},j}+h.c.)\\
 &-& J_{pl.} \sum_{{\bf x},j<k}(\hat{ U}_{{\bf x},j}\hat{ U}_{{\bf x}+\hat{j},k}\hat{ U}^\dagger _{{\bf x}+\hat{k},j}\hat{ U}^\dagger _{{\bf x},k}+h.c.).\cr\nonumber
 \end{eqnarray}
 We have used 
 \beq
\sum_{j=1}^{D-1}(e_{x,j}- e_{x-\hat{j},j})=n_{x,D},
\enq
to eliminate $n_{x,D}$. 
 Up to straightforward rescalings of the couplings, the first three terms are the same as in  1+1 dimensions \cite{prd92,prl121}, while the fourth one 
 requires at least one more spatial dimension and is as in the pure gauge case \cite{kogut79}. 
 Closely related derivations appear in Refs. \cite{judah.thesis,unmuth2018}. 
 The considerations \cite{meurice2019} regarding the modification of the algebra due to truncation and the relationship with the quantum link 
 approach \cite{CS96,brower97,qlink2,brower2020} remain valid. 
 
 The fact that we recover the Abelian version of the standard Kogut-Susskind (KS) form \cite{kogut79} follows from the usage of the 
 same time continuum limit scaling given by Eq. (\ref{eq:contlim}). However, in the original derivation, the operators 
 $\hat{U}_{x,\mu}$ are written as the exponentials of the spatial gauge field operators, while we have proceeded in a 
 gauge-invariant way by completely integrating over the gauge fields. Our derivation of the final algebraic result including Gauss's law shows that it follows exactly from the gauge-invariant Lagrangian definition. 
 
Notice that  Hamiltonians related to the KS Hamiltonian can appear in different contexts. For instance, 
Ref.  \cite{PhysRevX.5.011024} starts with a quantum many-body state for the matter fields which is invariant under a global symmetry. After introducing new degrees of freedom (the gauge fields), they construct a new state with a local symmetry. 
By combining this construction with the formalism of projected entangled-pair states (PEPS), they recover an Hamiltonian closely related to the KS Hamiltonian, in a framework that is convenient to explore low-parameter families of gauge-invariant states. Tensor network variational ansatzes for gauge-invariant states that can be connected to truncated KS 
models were also constructed in Ref. \cite{PhysRevX.4.041024}. In Refs. \cite{prl121,Unmuth-Yockey:2018ugm}, comparisons of the effects of truncations in the formalism used here  and variational methods were performed and it would certainly be useful to pursue this effort in new directions and to include the suggestions of Sec. \ref{subsec:robust}.

In this context, we would like to comment about the idea of considering a Hilbert space parametrized with 
new quantum numbers as shown in Eq. (\ref{eq:discrmax2}) and where all the states automatically satisfy Gauss's law. 
As the relation between the $e_{x,j}$ and $c_{x,j,k}$ is linear, we can study the effect of changing one of the $c_{x,j,k}$ by 
$\pm1$. For instance, $\Delta c_{x,1,2}=1$  generates the following changes: 
\beq
\Delta e_{x,1}=-1 ,\ \Delta e_{x+\hat{2},1}=1 ,\ \Delta e_{x,2}=1 ,\ \Delta e_{x+\hat{1},2}=-1.
\enq
This change can be visualized as an electric field circulating clockwise on a plaquette in the 1-2 plane and it clearly satisfies Gauss's law. The changes correspond to the $U^\dagger U^\dagger UU$ term in  the Hamiltonian (\ref{eq:ksham}).
For $D=3$, this is the end of the story and we can efficiently replace the term with two raising and two lowering operators by a term with a single raising or lowering operator \cite{unmuth2018}. 
The construction can be repeated for any pair of directions in higher dimensions, but as discussed in 
Sec. \ref{sec:transfer} the $c_{x,j,k}$ have some redundancy. For $D=4$, the 
geometric interpretation is easy with three spatial dimensions: we can combine 6 plaquettes on a cube in such a way that 
all the electric quantum numbers cancel. In other words, the effect of one of the $c_{x,j,k}$ can also be obtained with 
five others. For OBC, we could remove this redundancy by eliminating, for instance, all the $c_{x,2,3}$ except for those on a 2-3 plane at the boundary. For PBC, other sectors should be added in order to allow electric configurations wrapping around the spatial directions.

\section{Topological solutions and semi-classical approximations}
\label{sec:top}
In Sec. \ref{sec:eom}, we found a direct similarity between the continuous lattice equations of motion and the discrete 
tensor selection rules. In this section we discuss the effect of periodic boundary conditions on both sets of equations. 
We will limit ourselves to 
the solvable cases: the $D=1$ O(2) spin model and the $D=2$ pure gauge $U(1)$ model. 

For the $D=1$ O(2) spin model with PBC and $N_\tau$ sites, the equations of motion (\ref{eq:eomo2}) with $A_{x,\mu}=0$ are equivalent to the 
statement that $\sin(\varphi_{x+\hat{1}}-\varphi_x)$ takes the same value on every link. 
These equations have many solutions and we will focus our attention on the 
ones that can be interpreted as continuous topological solutions in the continuum limit for PBC. 
If we impose that $\varphi_{x+\hat{1}}-\varphi_x$ is a small constant, we can obtain a solution that meets 
this requirement. 
Given any choice for the constant, we can then ``integrate" the equations: starting with some $\varphi_0$, we obtain 
$\varphi_1$, and so on until, due to PBC, we get an independent value for $\varphi_0$ which should be consistent with the 
initial value modulo an integer multiple of $2\pi$. This approximately corresponds to a smooth mapping of the circle into itself 
provided that the successive changes can be made arbitrarily small. This can be accomplished by requiring that for all links
 \beq
\varphi_{x+\hat{1}}-\varphi_x=\frac{2\pi}{N_\tau}\ell, 
\enq 
for a given integer $\ell$. By taking, $N_\tau$ large with fixed $\ell$ we obtain a solution which can be interpreted as a topological solution 
with winding number $\ell$. In the limit $\ell \ll N_\tau$, these solutions have classical action
\beq
S_\ell\simeq \frac{\beta}{2}(\frac{2\pi}{N_\tau}\ell)^2 N_\tau.
\enq

We can calculate the quadratic fluctuations with respect to this solution. 
We can first use the global O(2) symmetry to set $\varphi_0=0$. Other values of $\varphi_0$ are taken into account 
by performing the integration over $\varphi_0$ which with our normalization of the measure yields a factor 1. 
By construction, the linear fluctuations vanish because the first derivatives are zero and all we need to calculate are the quadratic fluctuations
\beq
\Delta=\prod_{x=1}^{N_\tau-1}\int_{-\pi}^{\pi}\frac{d\varphi _x}{2\pi} e^{-S_\ell^{quad.}},
\enq
with 
\beq
S_\ell^{quad.}=\frac{\beta}{2} \cos(\frac{2\pi}{N_\tau}\ell)( \varphi_1^2+( \varphi_2-\varphi_1)^2+\dots +\varphi_{N_\tau-1}^2)
\enq
Following the standard quadratic path integral procedure, we find
\beq
\Delta=N_\tau^{-1/2} (2\pi\beta \cos(\frac{2\pi}{N_\tau}\ell) )^{-(N_\tau-1)/2}.
\enq

We can now attempt to re-sum the topological contributions. This is delicate because we have assumed $\ell \ll N_\tau$, however if $\beta$ is large 
enough, the terms with large $\ell$ are exponentially suppressed. In the same spirit, we will ignore the $\ell$ dependence of $\Delta$ and use the Poisson 
summation formula
\beq
\sum_{\ell =-\infty}^{\infty} e^{-\frac{B}{2}\ell^2}=\sqrt{\frac{2\pi}{B}}\sum_{n =-\infty}^{\infty} e^{-\frac{(2\pi)^2}{2B}n^2},
\enq
with $B=\beta (2\pi)^2/N_\tau$. Putting everything together, we get a semi-classical approximation of the 
partition function in the large $\beta$ limit
\beq
\label{eq:sc}
Z\simeq (2\pi\beta)^{-N_\tau/2}\sum_{n =-\infty}^{\infty} (e^{-\frac{n^2}{2\beta}})^{N_\tau}.
\enq

We now consider the solutions of the discrete Eq. (\ref{eq:noether}). The solution is that $n_{x,1}$ should be constant. 
With PBC, this implies 
the exact expression:
\beq
\label{eq:o2exact}
Z=\sum_{n =-\infty}^{\infty} (e^{-\beta}I_n(\beta))^{N_\tau}, 
\enq
which can be compared to the semi-classical expression Eq. (\ref{eq:sc}). 
Using the large $\beta$ approximations 
\beq
e^{-\beta}I_0(\beta)\simeq \frac{1}{\sqrt{2\pi \beta}}(1+ \mathcal{O}(1/\beta)),
\enq
and Eq. (\ref{eq:tn}) in the same limit, 
we see the approximate correspondence between the two expressions.

A similar construction can be carried for the $D=2$ pure gauge $U(1)$ model with PBC. We consider a rectangular $N_s \times N_\tau $ lattice.
The equation of motion requires that $\sin(f_{x,1,2})$ is constant. 
Following the analogy with the O(2) case, we start with
 \beq
 f_{x,1,2}\equiv A_{x,1}+A_{x+\hat{1},2}-A_{x+\hat{2}, 1}-A_{x,2}=\delta ,  \enq
 with $\delta$ a constant to be determined with PBC. 
As seen in Sec. \ref{sec:gf}, we can gauge fix the temporal links with a given spatial coordinate $x_1$ to the identity with the exception of one time layer. 
For definiteness, we take this layer of nontrivial time links to be between $\tau=N_\tau-1$ and $N_\tau$ which is identified with 0 due to PBC. 
The space links with a given spatial coordinate, which can be visualized as a vertical ladder can be treated as the indices of a $D=1$ O(2) model changing by $-\delta$ at each step until 
we get to the ``last" rung and temporal links are present. The constancy of the ``last" plaquette requires that 
\beq
A_{(x_1+1,N_\tau-1),2}-A_{(x_1,N_\tau-1),2}=N_\tau \delta.
\enq
Iterating in the spatial direction, we obtain PBC in the spatial direction provided that  
\beq \delta =\frac{2\pi}{N_s N_\tau}\ell.\enq
The action for this topological solution is 
\beq
S^{U(1)}_\ell\simeq \frac{\beta}{2}(\frac{2\pi}{N_s N_\tau}\ell)^2 N_s N_\tau.
\enq
Note that we could have obtained another periodic solution by setting {\it all} the 
time links to 1 and imposing PBC in time for $N_s$ independent $D=1$ O(2) models, however, 
the action for these configurations is larger by a factor $N_s^2$. 

The quadratic fluctuations can be calculated as in the O(2) case but with extra complications due to the 
special time layer. Keeping track of all the $2\pi$ factors and using Poisson summation for the winding numbers, we obtain the 
semi-classical approximation
\beq
\label{eq:sc2}
Z^{U(1)}\simeq (2\pi\beta)^{-N_s N_\tau/2}\sum_{n =-\infty}^{\infty} (e^{-\frac{n^2}{2\beta}})^{N_s N_\tau},
\enq
which agrees with the exact expression at leading order. 

\begin{figure}[h]
 \centering
\includegraphics[width=8cm]{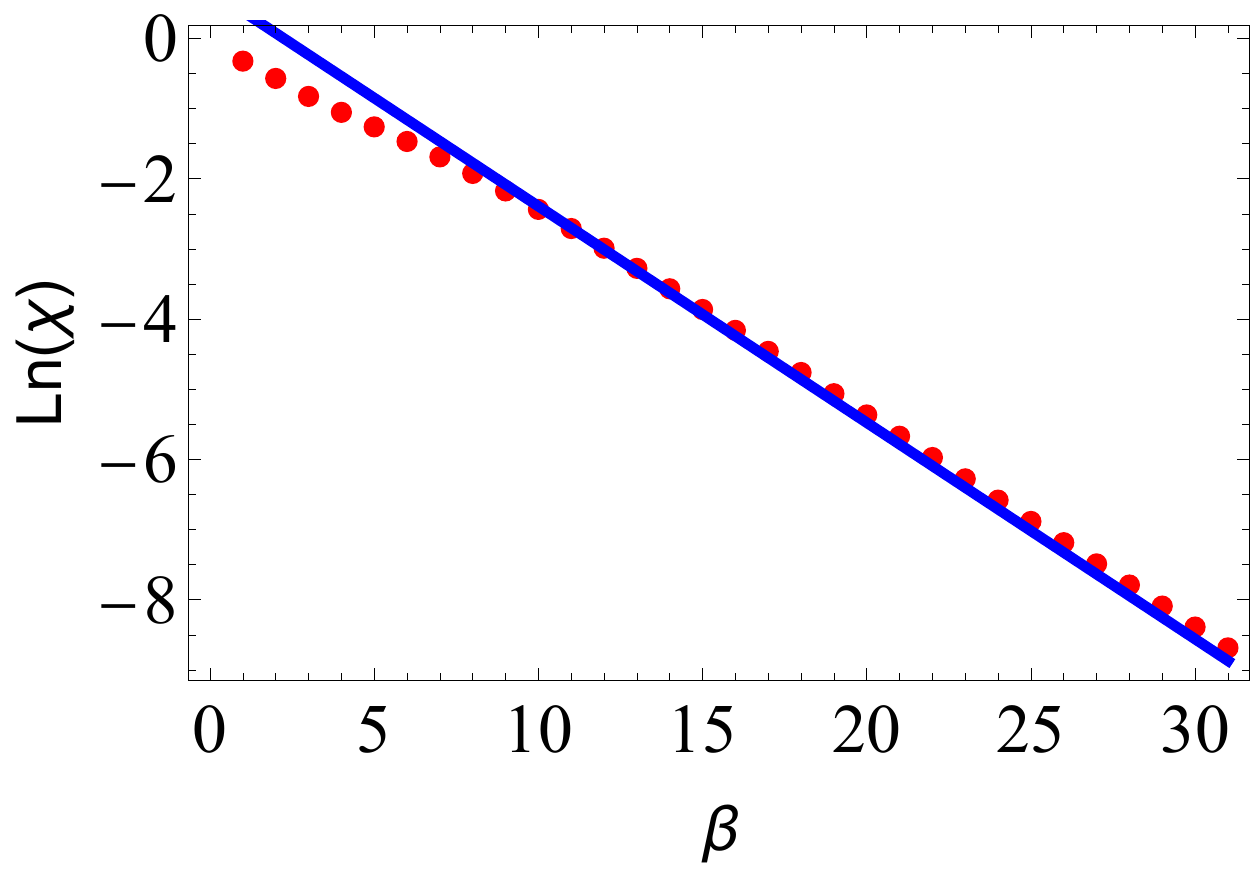}
\caption{\label{fig:susc} Logarithm of the topological susceptibility using the exact formula for $N_s=N_\tau=8$ expanded up to order 5 (dots) and the semi-classical approximation 
Eq. (\ref{eq:scapp}) (continuous line).} 
\end{figure}

As a test of the semi-classical picture we can calculate the topological susceptibility.
For this purpose we first calculate
\beq
Z(\beta, \theta)=\prod_{x,\mu}\int_{-\pi}^{\pi}\frac{dA_{x,\mu}}{2\pi} e^{-S_{gauge}-i\theta Q},
\enq
with the topological charge $Q$ defined as 
\beq
Q=\frac{1}{2\pi}\sum_{x} \sin(A_{x,1}+A_{x+\hat{1},2}-A_{x+\hat{2}, 1}-A_{x,2}).
\enq
The topological susceptibility is defined as 
\beq
\chi=-\frac{d^2}{d\theta^2}\ln(Z)|_{\theta=0}.
\enq
It can be calculated using the exact resummation \cite{gattringer2015} 
\begin{eqnarray}
Z(\beta, \theta)=\sum_{n =-\infty}^{\infty} &[&e^{-\beta}I_n(\sqrt{\beta^2-(\frac{\theta}{2\pi})^2}) \cr
&&\times ( \frac{\beta-\frac{\theta}{2\pi}}{\beta+\frac{\theta}{2\pi}})^{n/2}]^{N_s N_\tau}.
\end{eqnarray}
If $\chi$ is dominated by configurations corresponding to winding number $\pm 1$ where 
$|Q|\simeq 1$ in the 
continuum limit, we have the large-$\beta$ estimate
\begin{eqnarray}
\label{eq:scapp}
\chi \simeq &&(0)^2 1+(1)^2\exp(-\frac{\beta}{2}\frac{(2\pi)^2 (1)^2}{N_s N_\tau})\cr
&+&(-1)^2\exp(-\frac{\beta}{2}\frac{(2\pi)^2}{N_s N_\tau}(-1)^2).
\end{eqnarray}
Fig. \ref{fig:susc} shows that this estimate is reasonably good when $\beta$ is large enough.  

As a remark, it is a common misconception to identify the Fourier mode indices $n$ in Eq. (\ref{eq:o2exact}) as ``topological sectors". They are rather labelling ``rotor energy levels" 
$n^2/2$. The fact that Poisson summation interchanges these energy levels with the correctly 
identified topological sectors was observed in Ref. \cite{akerlund2015} in a version of the O(2) model where the fluctuations are limited. Note also that it is possible to construct models where the approximations Eqs. (\ref{eq:sc}) and (\ref{eq:sc2}) are exact. The questions of topological 
configurations and duality are discussed for Abelian gauge models of this type in various dimensions in Refs. \cite{gattringer2018,sul2019}. 

\section{Conclusions}

In summary, we have shown that some standard theorems and identities associated with the $U(1)$ symmetry that can be derived in the conventional formulation of field theory have a discrete counterpart in TFT. This includes the equations of motion, Noether's theorem, Maxwell's equations with charges and currents, Gauss' law, gauge-fixing and effects of boundary conditions.

We have constructed gauge-invariant transfer matrices by reorganizing the partition function obtained integrating over all the fields without gauge fixing.
We also explained how an equivalent partition function is obtained by a gauge-fixing which removes intermediate 
projections into the sector of the Hilbert space which satisfies Gauss's law. These projections are only useful 
in the case of a noisy evolution. We proposed a reparametrization of the the sub Hilbert space satisfying Gauss's law 
in arbitrary dimension which generalizes dual construction in $D=3$ \cite{unmuth2018}.

Practical implementations of TFT require finite truncations.  They provide numerically accurate 
approximations at finite volume \cite{pra90,prd92,prl121,butt2019}. The results derived here depend only on the 
selection rules which completely capture the symmetry and not on the numerical values of the Bessel functions appearing 
in Fourier expansions. This confirms the observation that truncations preserve the symmetries \cite{meurice2019}.
The class of universality is encoded in the selection rules of the tensors and 
it is expected that in the continuum limit, results should not depend on microscopic details. 
Similar expectations are found in the quantum link approach \cite{CS96,brower97,qlink2,brower2020}. One advantage of TFT is that it connects smoothly the 
Lagrangian and Hamiltonian approaches in a way that allows testing using standard importance sampling methods. This allows comparisons with Hamiltonian based quantum simulations proposals for Abelian gauge models \cite{Zohar:2012ay,
Tagliacozzo:2012vg,
Kuno:2014npa,
Kasamatsu:2012im,
Celi:2019lqy,
Surace:2019dtp}.

The discrete nature of TFT formulations makes it a generic tool to setup quantum computing protocols. It provides an alternative to field discretization \cite{Hackett:2018cel,Alexandru:2019nsa,Lamm:2019bik}. 
Motivations for quantum computing include doing ab-initio  real-time calculations relevant to fragmentation processes and parton distribution functions \cite{lamm2019}. In order to work on these ambitious and high-impact projects, we need to move up the steps of a ``ladder" of models \cite{kogut79,kogut83} that has been proven effective to deal with the static properties of hadrons. 
The first steps are the spin and gauge Ising models. Practical implementations are discussed in Refs. \cite{lamm2018, gustafson2019a,gustafson2019b}. The next steps are their counterparts with a continuous and compact Abelian $U(1)$ symmetry 
\cite{unmuth2018,kaplan2018,PhysRevA.99.042301,brower2020}. 
In this context, the Euclidean transfer matrix could also be used to prepare initial states following the suggestion of Ref. \cite{harmalkar2020}. 

Models with Wilson \cite{Shimizu:2014fsa,Shimizu:2014fsa,Takeda:2014vwa,Shimizu:2017onf,Nakamura:2018enp,
Kadoh:2018hqq,Kadoh:2018tis} and staggered \cite{butt2019} fermions have also been reformulated using TFT. The Schwinger model is of great interest in this context. This model and its \zq approximations have been studied directly with the Hamiltonian formalism \cite{Byrnes:2002nv,Banuls:2013jaa,Kuhn:2014rha,Buyens:2015tea,PhysRevD.98.074503,magnifico2019}, providing useful comparisons for future TFT calculations.

\begin{acknowledgments}
We thank the late D. Speiser and C. Itzykson for their teaching on Pontryagin duality, Peter-Weyl theorem and strong coupling expansions. 
 We thank J. Unmuth-Yockey for many discussions on Abelian gauge theories. 
 We thank R. Edwards and JLab for the invitation to give lectures that helped sharpen some of the arguments made in the paper. 
 We thanks M. Vander Linden for help with figures.  
 This work was supported in part by the U.S. Department of Energy (DOE) under Award Numbers DE-SC0010113, and DE-SC0019139.
\end{acknowledgments}

 
%

\end{document}